\documentclass[12pt,showpacs,preprintnumbers]{revtex4}

\usepackage{graphicx}
\usepackage{dcolumn}
\usepackage{bm}
\usepackage{amssymb}
\usepackage{epsfig}    
\def\beq{\begin{equation}}
\def\eeq{\end{equation}}
\def\ba{\begin{array}}
\def\ea{\end{array}}
\def\bea{\begin{eqnarray}}
\def\eea{\end{eqnarray}}

\def\End{\end{document}}

\begin{document}                                                              

\title{Generalized  Lorentz Transformations}

\author{Virendra Gupta}
\email{virendra@mda.cinvestav.mx}

\affiliation{
\vspace*{2mm} 
Departamento de F\'{\i}sica Aplicada, 
CINVESTAV-M\'erida, A.P. 73, 97310 M\'erida, Yucat\'an, M\'exico}

\begin{abstract}
   Generalized Lorentz transformations with modified velocity parameter
    are considered. Lorentz transformations depending on
   the mass of the observer  are suggested.The modified formula for 
   the addition of velocities remarkably preserves the constancy of the 
   velocity of light for all observers. The Doppler red shift is affected 
   and  can provide a test of such generalisations.
\end{abstract}

\maketitle


\section{Introduction}
     
     Over the years,modification and generalization of  Lorentz transformations has been considered by many 
     authors.Specifically, addition of Lorentz invariance violating interactions to the Standard Model has
      been considered  \cite{CK,CG}.Also,extended linear and non-linear Lorentz transformations  have been 
      considered \cite{DF}.These papers contain many references to other work in this area.

     The usual Lorentz transformations involve the dimensionless velocity parameter
       $\beta= v/c$  and the dilatation factor $\gamma=\frac{1}{\sqrt{1-\beta^2}}$.
    The important point to note is that with a general velocity parameter $ B$ (a function of $\beta= v/c$
    and possibly other parameters) Lorentz invariance is guaranteed as long as the corresponding
    dilatation factor  $ G$ is  the same function of the new velocity parameter $B$
    as the old $\gamma$ was of $\beta= v/c$. In other words,$G=\frac{1}{\sqrt{1-B^2}}$.
   
    An important constraint on any  velocity parameter is that it be equal to zero for $v=0$
     and be equal to unity for $v=c$. The latter constraint guarantees  Einstein's second
     postulate, namely  that the velocity of light $c$ is the same for all observers.
     
First we explore the possibility of more general Lorentz transformations with $\beta$ replaced by 
the mass dependent velocity parameter $\beta(m)= v/c(m)$.
Section 2 contains the definition of $ c(m)$ consistent with the present experimental constraints.
Section 3 gives the modified law of addition of velocities. It is pointed out that this
formula still gives that the velocity of light is the same for all observers. In Section 4,
it is suggested that accurate  measurements of the Doppler shift could provide a possible test 
of the mass dependent  Lorentz transformations considered here.
  In section 5, Lorentz transformations depending on a general velocity parameter $B$ which is
  a function of only $\beta= v/c$ are considered.
\section{Choice of $c(m)$}.
              
The mass dependent velocity  $c(m)$ is defined to be 
\beq
           c(m)=c[1+ F(\zeta)],
\eeq
where $F$ is an analytic function of the dimensionless variable $\zeta$ defined as 
$\zeta =m/ P_M$, where the Planck mass\cite{PDG}
$P_M\approx 1.22\times 10^{19}Gev/c^2 \approx 2.18\times10^{-5}$ gr.
                  
 To conform with present knowledge the function $F$ must satisfy the following constraints:
 A) For $m=0$, that is $\zeta =0$ , $F(0)=0$. This implies that $c(0)=c$.
 B)For sub-atomic particles (electron ,proton etc.) $\zeta $ is very small. Since,our accelerators work,
 this implies that $F(\zeta)$ must be negligibely small for small $\zeta$. C) Our understanding of
planetary motion, constrains  $F(\zeta)$ to be negligibely  small for large $\zeta$.
  With these constraints in mind we had earlier considered \cite{VG}
   \beq
                       F(\zeta)=\zeta^n exp(-\zeta^n)
   \eeq
where  $n$ is a positive real number. This function has a maximum value $e^{-1}=0.367879$
for  $\zeta=1$ independent of the value of $n$. It is centered around $\zeta=1$. As $n$
increases, the height remains the same, but it becomes narrower and narrower. For extremely
large $n$ (tending to infinity) this sequence of functions tends to  a vertical  line of height
$e^{-1}$ at $\zeta=1$. In the limit it is like a ``finite Dirac delta-function". A plot of the function
is given in reference\cite{VG}.
              
\section{Lorentz transformations depending on mass}
        
Consider parallel Cartesian coordinate systems, $S$ and $S'$ whose origin coincided at $t=t'=0$.
Let their relative velocity be $v$ along the x-axis as measured by $S$ . Then
\beq          
ct'=\gamma (ct-\beta x), \quad
x'=\gamma(x-\beta ct),\quad
y'=y, \quad z=z',
\eeq
where
\beq 
                  \beta=\frac{v}{c},\quad    \gamma=\frac{1}{\sqrt{1-\beta^2}}.
\eeq
This is the usual Lorentz transformation. It leaves the space-time interval invariant,
that is $ S^2=(ct)^2 -r^2=(ct')^2-r'^2=S'^2$ where $ r^2=x^2+y^2+z^2$.
The crucial point is that the invariance of $S^2$ is \emph{guaranteed} for any $\beta$
as long as $\gamma $ is defined  in terms of  $\beta$ as above in Eq (4).
Thus, if we replace the velocity parameter $\beta$ by the mass dependent velocity parameter
\beq
                      \beta(m)=\frac{v}{c(m)}
\eeq
and $\gamma$ by
\beq             
                     \gamma(m) =\frac{1}{\sqrt{1-\beta(m)^2}}.
\eeq
in Eq.(4), then the space-time interval will be invariant  under these mass dependent Lorentz
transformations.

In the real world the observer and the observed have masses and the mass dependent
Lorentz transformations defined by $\beta(m)$ may be relevant.The mass which enters here is
the mass of the observer in frame $S'$.
The mass dependent velocity parameter will give a modified relative velocity formula.
For simplicity,consider three systems$ S_1,S_2$ and $ S_3 $ with aligned $x$-axes. Let $ S_2$ be moving 
along the $x$-direction with velocity $v$ respect to $ S_1$ and let $ S_3 $ be moving with velocity
$v'$ along the $x$-direction with respect to $ S_2$. Let the corresponding mass dependent velocity 
parameters be
\beq
                \beta(m) =\frac{v}{c(m)}\quad\mbox{and}\quad \beta(m')=\frac{v '}{c(m')}.
\eeq
Then the relative speed of $ S_3 $ with respect to $ S_1$ will be
\beq
             \beta''(m,m') = \frac{\beta(m) +\beta(m')}{1+\beta(m)\beta(m')}.
\eeq  
This mass-dependent relativistic formula for addition of velocities reduces to the
usual formula with the replacements $\beta(m)=\beta$ and  $\beta(m')=\beta' $ .
Note that , since for $m=0 $(photons),  $\beta(0)=\beta=1$ ,the relative velocity $\beta''=1$ !.
Also, if $m=m'=0$ then $\beta''=1$ ! In other words,the mass-dependent relative velocity
formula above respects Einstein's postulate that the velocity of light is same for all 
observers!

\section{Doppler shift}

An experimental test of the mass-dependent Lorentz transformations considered here can 
come from extremely accurate measurements of the Doppler shift, especially for objects  with masses
 of the order of the Planck mass. For the frames $S$ and $S'$ moving with relative velocity
$v$ in the $x$ direction  (considered above), the energy( in $S'$ ), namely
\beq             
                   E'=\gamma(m)( E- \beta(m)cp_x).
\eeq        
In terms of the frequencies $\nu$ and $\nu'$, this gives
\beq 
              \nu'= \gamma(m) (1 - \frac{v}{c})\nu.
 \eeq 
This is the usual formula with $\gamma$ replaced by $\gamma(m)$. Since, $F(\zeta)$ is chosen to be
extremely  small except  for  masses of the order of the Planck mass ($\zeta\approx1$) and since $m$
 is the mass of the observer, different observers with different masses should see different red shift
 from the same source!
 However,this will not happen for mass dependent Lorentz   transformations
 which are symmetrical in the masses $m$ and $ m'$ in the  frames $S$ and $ S'$.
In this case,we take the  the mass-dependent velocity parameter to be
\beq
    B(m,m')= v/c(m,m'),
\eeq
where
\beq
       c(m,m')=c[1+\sqrt{F\zeta)F(\zeta')}].
 \eeq
  The corresponding dilatation parameter will be
  \beq
       G(m,m')=\frac{1}{\sqrt{1-B(m,m')^2}}. 
 \eeq
 Since, $B(m,0)=B(0,m')=B(0,0)=\beta$ the velocity of light will be the same for all observers

 One needs new experiments particularly for observers with masses in the
 Planck mass range .

\section {Generalized   Lorentz   transformations   depending  only  on the velocity 
         parameter  $\beta$ }

       In this case, the general velocity parameter $B$ is  a function of $\beta$ only. Clearly,
     there are many posssible choices for $B(\beta)$ which satisfy $B(0)=0$ and $B(1)=1$.

     A possible choice is that $B(\beta)$ is a curve passing through the points $(0,0)$ and
     $(1,1)$ in the $\beta$-$B$ plane. An obvious choice is $B= Sin(\frac{\pi\beta}{2})$.
     For small velocities  this is approximately equal to $\beta$. However, for $\beta =0.5$
     its value is approximately equal to $0.7$.

    Another  simple possibility is that the curve is a circle in the $\beta-B$ plane
    passing through the points  $(0,0)$ and $(1,1)$, so that  the part of the straight 
     line $B=\beta$ between the points  $(0,0)$ and $(1,1)$ is a chord. The equation
     of such a circle is
 \beq 
        [\beta -A]^2 + [B-(1-A)]^2 = A^2 +(1-A)^2 .
\eeq 
      Larger the radius of the circle,that is larger the value of $A$, more closely the circle 
     will approximate the straight line $B=\beta$ between the points $(0,0)$ and $(1,1)$.
      Quantitatively, for large $A>0$, the centre of the circle will be at $(A,-A)$ in the fourth
      quadrant and the usual midway point  $(\beta,B)=(0.5,0.5)$ is replaced by the point
      $(\beta,B)=(0.5 -(4A)^{-1},0.5+(4A)^{-1}$.Thus,such a circle cwould lie above but very close
      to the usual line $\beta$=$B$ between the points $(0,0)$ and $(1,1)$. An alternative 
       possibility is  obtained by changing $A$ to $-A$.Such a a circle will have its centre
       in the second quadrant at $(-A,A)$ and lie below but very close to the usual line $\beta$=$B$
       between the points $(0,0)$ and $(1,1)$.
        Since,the usual $\gamma$ is replaced by $G=\frac{1}{\sqrt{1-B^2}}$,
        again the Doppler shift will be affected and would provide tests of such generalizations.
 
\section{Concluding remarks}
 We have considered generalized  Lorentz transformations such that  the velocity of light d
  is the same for all observers.
 It is pointed out that detailed measurements of  the Dppler effect can provide a test of the
 generalizations presented here. Experiments are needed  over a large range of velocities  to
  check that the velocity parameter is indeed $\beta=v/c$.

\noindent
{\bf Acknowledgements}~~~~~I am very grateful to  Antonio Bouzas and Francisco Larios for discussions and
help with the manucscript.I would also like to thank CONACYT and SNI for support.

\end{document}